\documentclass[useAMS,usenatbib,usegraphicx]{mn2e}

\title[Emission-line Stars in M33's Giant H~II Regions]{Wolf-Rayet Stars in M33 II: Optical Spectroscopy of emission-line stars in Giant H{\sc ii} 
Regions}

\author[L. Drissen et al.]{Laurent Drissen$^1$\thanks{E-mail: ldrissen@phy.ulaval.ca},
Paul A. Crowther$^{2}$,
Leonardo \'Ubeda$^1$,
and Pierre Martin$^3$\\
$^1$ D\'epartement de physique, de g\'enie physique et d'optique, Universit\'e Laval, Qu\'ebec, G1K 7P4, Canada;\\
$^2$ Department of Physics and Astronomy, University of Sheffield, Hounsfield Road, Sheffield S37RH;\\
$^3$ Canada-France-Hawaii Telescope, 65-1238 Mamalahoa Hwy, HI 96743, USA}

\begin{document}
\date{Accepted ---. Received ---; in original form 2008 April 4}
\label{firstpage}

\maketitle

\begin{abstract}
We present optical spectra of 14 emission-line stars in M33's giant
HII regions NGC 592, NGC 595 and NGC 604: five of them are
known WR stars, for which we present a better quality
spectrogram, eight were WR candidates based on narrow-band
imagery and one is a serendipitous discovery.
Spectroscopy confirms the power of interference filter imagery to detect
 emission-line stars down to an equivalent width of about 5
\AA\ in crowded fields.
We have also used archival HST/WFPC2 images to correctly identify emission-line stars in NGC 592 and NGC 588.
\end{abstract}


\section{Introduction}
The overwhelming majority of stars in the Universe display
absorption lines in their visible range spectrum.
Emission lines in stellar spectra are therefore notable exceptions
which betray fascinating properties such as the presence of
unusually strong chromospheric activity, rapid rotation or in
the case of interest to this paper, strong mass loss and high
luminosity. Nearby giant H~{\sc ii} regions are hosts to an
interesting zoo of emission-line objects (Walborn \& Fitzpatrick
2000), most of them being post-main sequence massive stars
of notable interest for our understanding of stellar evolution
at the top end of the initial mass function. They also offer
important testbeds to understand the more distant, 
unresolved starbursts: individual stars can be counted and 
spectroscopically classified, allowing
a direct comparison with the modelisation of the spatially integrated
properties of their ionizing cluster (see Vacca et al. 1995 
for such a comparison in 30 Doradus, or Bruhweiler, Miskey \&
Smith Neubig 2003 in NGC 604). 
Among massive stars with emission lines, those of Wolf-Rayet (WR) 
type are the easiest to detect and
classify because of their strong and broad emission lines in the 
visible part of the spectrum. Thanks to surveys in Local Group 
galaxies (Massey \& Johnson 1998)
and improvements in theoretical models (Meynet \& Maeder 2005), 
the evolutionary status of WR stars is now understood well enough 
to use them as diagnostics to infer the properties of starburst 
regions. For instance, in unresolved clusters or starburst knots 
of distant galaxies, the equivalent width of the
`WR-bumps' are good indicators of the age and upper mass limit of the
stellar population (Pindao et al. 2002).

The small spiral galaxy M33 is host to four giant H{\sc ii} regions 
bright enough to have their own NGC number: NGC 604, the second most 
luminous starburst cluster in the Local Group; then, in decrasing 
order of H$\alpha$ luminosity, NGC 595, NGC 592 and NGC 588. Despite their different galactocentric distances, these four regions have very similar metallicities, with $12 +$ log O/H $= 8.4 - 8.5$ (Magrini et al. 2007).
Two papers published in the fall of 1981 presented 
the spectroscopic discovery of WR stars in these 
clusters: D'Odorico \& Rosa (1981) derived a
surprisingly large (50) number of WR stars in NGC 604, 
while Conti \& Massey
(1981) noticed that some WR stars in the four regions
were excedingly luminous. Both studies however suffered from a 
lack of spatial resolution. More WR candidates were identified 
by interference filter imagery (taking advantage of their strong
HeII emission) and spectroscopically confirmed by 
Massey \& Conti (1983), Armandroff \& Massey (1985, 1991) and Massey, 
Conti \& Armandroff (1987). The most detailed catalogue of WR
stars (with spectral classification) and WR candidates in 
M33 is presented in Massey \& Johnson (1998).
Drissen, Moffat \& Shara (1990, 1993) identified more WR candidates based on high 
resolution CCD images with interference filters; but until now however, 
none of these were spectroscopically confirmed.

In the first paper (Abbott et al. 2004) of this series dedicated
to WR stars in M33, we presented new spectra of one Of, 14 WN,
one transition-type WN/WC and 26 WC stars in the field of M33.
In this second paper, we present spectra of known WR stars and
most of the WR candidates in the giant HII regions NGC 604, NGC
595 and NGC 592. 

\section{Observations}
The spectroscopic data were obtained with the 
Multi-Object Spectrograph (MOS) attached
to the 3.6-m Canada-France-Hawaii Telescope (CFHT) in 2000 
September and 2001 October. Prior to spectroscopy, a high
quality image of each region was obtained in order to identify
the targets and position the slitlets with sub-acrsecond
accuracy. The seeing during both observing runs ranged from 0.7$''$ to 1.0$''$, except when the observations of NGC 592 were obtained (1.5$''$ at high air mass; see below). After the spectra were
obtained, a superposition of these images with the resulting 2D
spectral images ensured a correct {\it a-posteriori} identification of the stars. 
We used the B600 grating, which, 
combined with the slit width of 1.5 arcsec, provided a spectral 
resolution of $\sim 9$ \AA . 
Exposure times were 900s for NGC 592, 2700s for NGC 604 and 2700s
for NGC 595. In the case of NGC 592, the observations were 
obtained at the end of the night with a long and wide (5$''$) 
slit, which reduced the spectral resolution. 
The data were then reduced using standard procedures in IRAF.

To complement these spectroscopic observations,  we have used archival images obtained with the
Wide Field and Planetary Camera 2 (WFPC2) and the Advanced Camera for Surveys (ACS) on board the {\it Hubble Space Telescope} to
correctly identify emission-line stars in all regions.Table 1 lists
the names and properties of the imaging datasets that we have used
for this research. The data were extracted from the Canadian Astronomy Data Centre's web interface.

\begin{table*}
\begin{minipage}{170mm}
\caption{Archival HST images used in this study    \label{tbl01}}
\begin{tabular}{@{}lccc@{}} 
\hline
 Region &  Program ID  & Dataset &   Exp. time \\ 
 Filter &    & (CADC)  & (sec) \\ 
\hline
NGC 588: & & & \\
F170W  & 5384            & U2C60701B  & 700  \\
F336W   & "    &  U2C60703B   &  320 \\
F439W   & "    &  U2C60801B   &  360 \\
F547M   & "     &  U2C60803B   &  200 \\
F469N   & "     &  U2C60805B   &  600 \\
 & & & \\
NGC 592: & & & \\
F170W  &   9127    & U6DJ010DB & 1560  \\
F255W  & "    & U6DJ0109B   &  1440  \\
F336W   & "    &  U6DJ0107B   &  520 \\
F439W   & " &  U6DJ0103B   &  600 \\
F555W   & "       &  U6DJ0101B  &  320 \\
 & & & \\
NGC 595: & & & \\
F170W  & 5384            & U2C60901B  & 700  \\
F336W   & "    &  U2C60903B   &  320 \\
F439W   & "    &  U2C60A01B   &  360 \\
F547M   & "     &  U2C60A03B   &  200 \\
 & & & \\
NGC 604: & & & \\
F336W   &  5237   &  U2AB0207B   &  1200 \\
F555W   &  "   &  U2AB0201B   &  400 \\
F547M   &  5773   &  U2LX0307B   &  1000 \\
F656N   &  " &  U2LX0301B   &  2200 \\
F673N   & "       &  U2LX0303B  &  2200 \\
F220W & 10419 & J96Y11010 & 600 \\
F250W & " & J96Y11020 & 800 \\

\hline
\end{tabular}  
\end{minipage}
\end{table*}

\section{Results}

Spectrograms of Of and WN stars in the spectral range of the strongest emission lines are shown in Figure 1, while those of the three WC stars are shown in Figure 2. Table~\ref{tbl02} summarizes the photometric and spectroscopic properties of  all the emission-line stars detected in these regions, from this and from previous work.
The equivalent widths listed in this table as EW (WR) refer to the entire emission lines in the ''WR bump'' 
between 4630 and 4700 \AA\ .

\begin{table*}
 \centering
 \begin{minipage}{170mm}
  \caption{Properties of emission-line stars in M33's giant HII regions.\label{tbl02}}
  \begin{tabular}{@{}lcccccrrll@{}}
  \hline
   Star\footnote{Newly confirmed emission-line stars appear in bold characters.}    &     RA  & Dec & B  & M$_{B}$\footnote{Indicative only: the absolute magnitude was determined assuming a distance of 850 kpc (distance modulus = 24.6) and a uniform extinction A$_B$ = 0.5 mag.} & EW(WR)\footnote{Equivalent width, in \AA\ ,  of the emission lines in the 4600 - 4700 \AA\ region.} &Sp. (old) & Sp (new) & Alt. names\footnote{CM: Conti \& Massey 1981; MC: Massey \& Conti 1983; MCA: Massey, Conti \& Armandroff 1987; MJ: Massey \& Johnson 1998; AM: Armandroff \& Massey 1985, 1991.}\\
  \hline
N604 - WR1 & 01:34:32.37 & +30:47:00.9 & 17.7  & -7.4 & ----- &WCE & ----- & CM11, MC74, MJ-WR135 \\
N604 - WR2 & 01:34:32.50 & +30:47:00.3 & 17.1  & -8.0  & ----- &WN & ----- & CM11, MC74, MJ-WR135 \\
N604 - WR3 & 01:34:32.69 & +30:47:05.4 & 17.8  &  -7.3 & ----- &WN & ----- & CM12, MJ-WR136 \\
N604 - WR4 & 01:34:32.65 & +30:47:07.1 & 17.1  &  -8.0 & ----- &WN & ----- &  CM12, MJ-WR136\\ 
N604 - WR5 & 01:34:32.83 & +30:47:04.6 & 19.0 &  -6.1 & 450 &WC & WC6 & CM12  \\
N604 - WR6 & 01:34:33.68 & +30:47:05.8 & 17.9 &  -7.2 & 35 &WN & WNL & CM13 \\
N604 - {\bf WR7} & 01:34:33.87 & +30:46:57.6 & 20.1 &  -5.0 & 160 &WR? & WC4 &  \\
N604 - {\bf WR8} & 01:34:32.17 & +30:47:07.0 & 18.5 & -6.6  & 18 &WR? & WN6 &  \\
N604 - {\bf WR10} & 01:34:32.74 & +30:46:56.5 & 18.5 &  -6.6 & 18 &WR? & WN6 &  \\
N604 - {\bf WR11} & 01:34:34.06 & +30:46:56.2 & 20.7 &  -4.4 & 6 &WR? & WNE &  \\
N604 - {\bf WR12} & 01:34:32.55 & +30:47:04.4 & 18.1 &  -7.0 & 15 &WR? & WN10 &  \\
N604 - {\bf WR13} & 01:34:35.15 & +30:47:05.8 & ----- &  ----- & 5 &----- & O6.5Iaf &  \\
N604 - {\bf V1} & 01:34:32.30 & +30:47:03.9 & 17.8 &  -7.3 & 13 &WR? & Of/WNL &  \\
 & & & & & & & & & \\
N595 - WR1 & 01:33:34.22 & +30:41:38.1 & 18.3  &  -6.8 & 47 &WNL & WN7 & CM6, MC32, AM6, MJ-WR49 \\
N595 - WR2a & 01:33:33.76 & +30:41:34.0 & 19.7 &  -5.4 & ----- &WNL? & ----- & CM5, MC31, MJ-WR47 \\
N595 - WR2b & 01:33:33.72 & +30:41:34.2 & 18.2 &  -6.9 & ----- &WNL? & ----- & CM5, MC31, MJ-WR47 \\
N595 - WR3 & 01:33:33.81 & +30:41:29.8  & 20.5 &  -4.6 & 300 &WC & WC6 &  MC29, AM5 \\
N595 - WR4 & 01:33:32.95 & +30:41:36.2  & 18.1 &  -7.0 & ----- &WN & ----- & AM4, MJ-WR43 \\
N595 - WR5 & 01:33:32.80 & +30:41:46.2  & 18.2 &  -6.9 & 80 &WNL & WN7h & AM3, MJ-WR42 \\
N595 - WR6 & 01:33:32.61 & +30:41:27.3  & 19.1 &  -6.0 & ----- &WN & ----- & MC28, AM2, MJ-WR41 \\
N595 - WR7 & 01:33:34.28 & +30:41:30.5  & 20.3 &  -4.8 & ----- & WN & ----- & AM7,MJ-WR51 \\
N595 - WR8 & 01:33:34.02 & +30:41:17.2  & 20.3 &  -4.8 & ----- &WN & ----- & MCA4, MJ-WR48 \\
N595 - {\bf WR9} & 01:33:33.28 & +30:41:29.8  & 20.5 &  -4.6 & 27 &WR? & WN7 &  \\
 & & & & & & & & & \\
N592 - WR1 & 01:33:11.81 & +30:38:53.1  & 18.1 &  -7.0 & ----- &WN & ----- & CM3,MC17,MJ-WR25 \\
N592 - {\bf WR2} & 01:33:12.42 & +30:38:48.4  & 19.7 &  -5.4& 25 &WR? & WN4 &  \\
N592 - MC19 & 01:33:15.00 & +30:38:07.4 & 19.8  &  -5.3 & ----- &WCE & ----- & MJ-WR30 \\
N592 - MCA2 & 01:33:10.70 & +30:39:00.6  & 20.8 &  -4.3 & ----- &WN & ----- & MJ-WR24  \\
 & & & & & & & & & \\
N588 - UIT008 & 01:32:45.33 & +30:38:58.4  & 17.6 & -7.5 & ----- &Ofpe/WN9 & ----- & MJ-WR5, J1 \\
N588 - MC3 & 01:32:45.66 & +30:38:54.4  & 18.1 &  -7.0 & ----- &WN & ----- & CM1, MC3,MJ-WR6,J2\\
\hline
\end{tabular}
\end{minipage}
\end{table*}

\begin{figure}
\includegraphics[width=84mm]{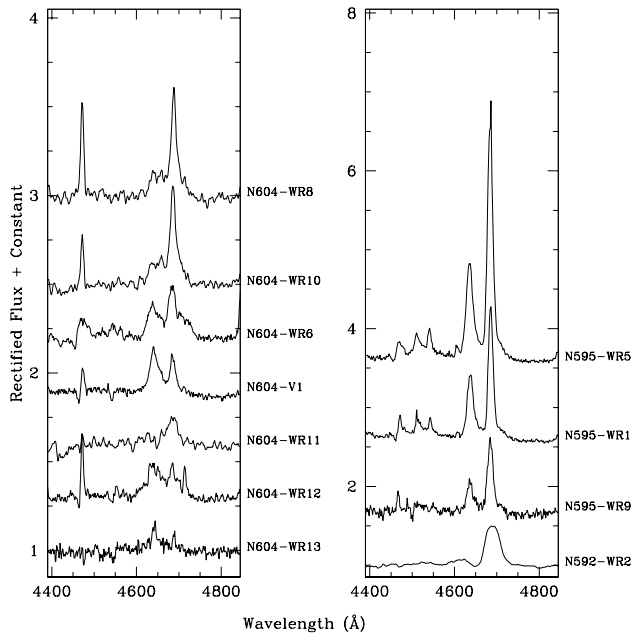}
\caption{Rectified spectra of Of and WN stars in our sample, in the 4600 \AA\  region. We should note that much of the strong, narrow emission line at 4471 \AA\  in the spectra of N604-WR8 and WR10 is of nebular origin.}
\end{figure}

\begin{figure}
\includegraphics[width=84mm]{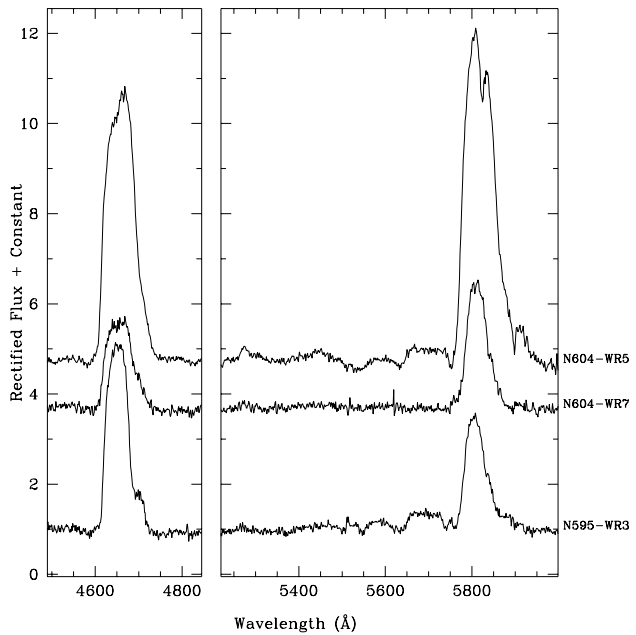}
\caption{Rectified spectra of WC stars in our sample, in the wavelength ranges of the most prominent lines.}
\end{figure}

\begin{figure}
\includegraphics[width=84mm]{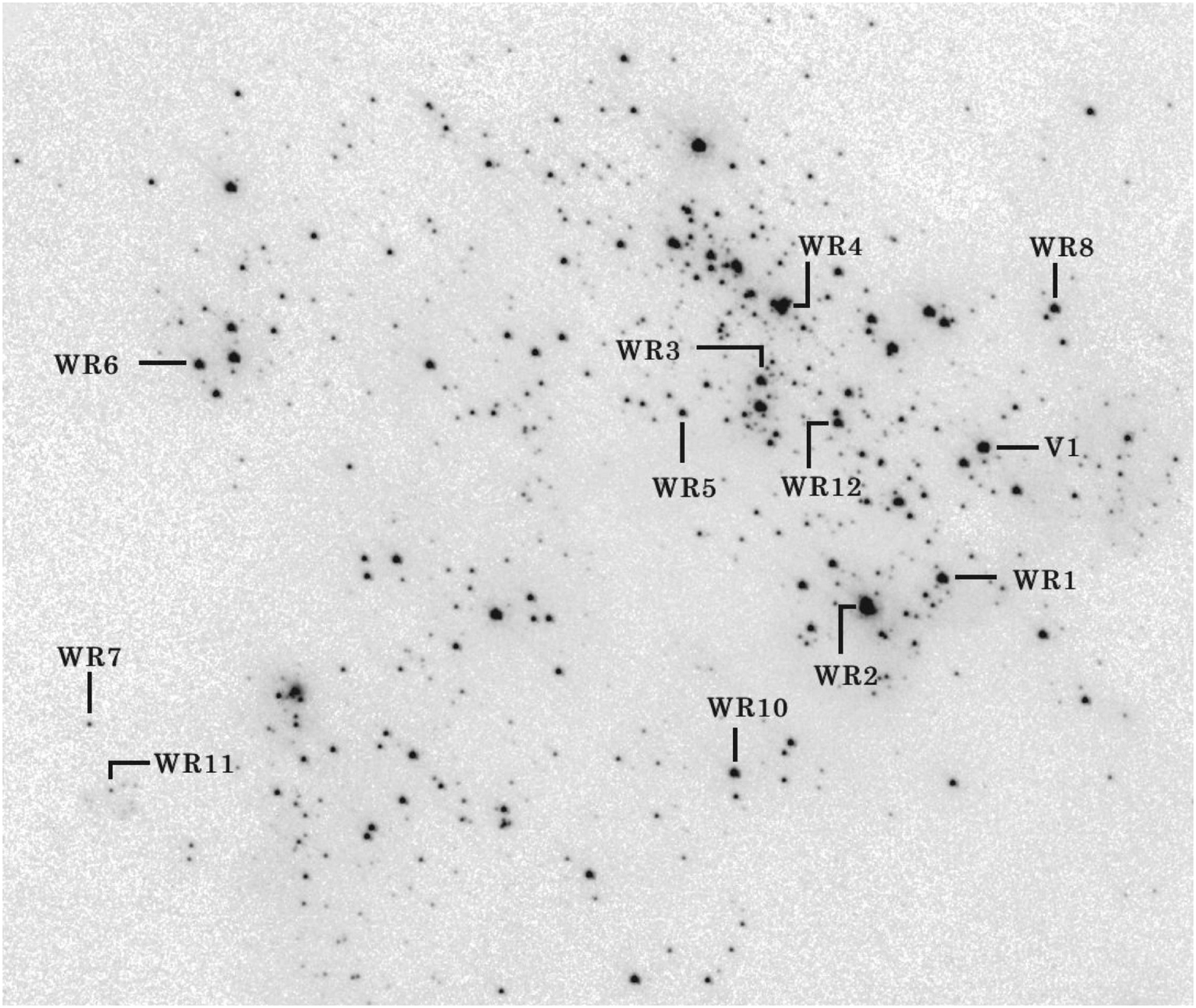}
\caption{Finding chart for emission-line stars in NGC 604, from an ultraviolet HST/ACS image (combination of F220W and F250W filters). Star WR13 is outside this field of view, and is shown in Figure 6. The field of view is 23$'' \times 20''$ (80 pc $\times 70$ pc), with North to the top and East to the left.}
\end{figure}

\subsection{NGC 604}

Drissen, Moffat \& Shara (1993; hereafter DMS93) presented finding charts for the W-R candidates in NGC 604, from aberrated
WF/PC images. We update this chart in Figure 3 with a much better quality image obtained from a series of archival HST/ACS images (filters F220W and F250W).

Bruhweiler et al. (2003) were able to derive the spectral type of 
40 stars in the central region of NGC 604 based on ultraviolet 
spectral imagery with HST's Imaging Spectrograph (STIS). UV
spectroscopic classification is not hampered by the very strong
nebular emission lines present in the visible region; this
survey is therefore the most complete stellar census so far 
in this cluster. We now describe the spectra of the newly observed stars.

WR5 was first identified as a WR by Conti \& Massey (1981), but
their spectrogram also included neighboring WR4 and WR3; it was
thus classified as a WN star. However, Drissen, Moffat \& Shara (1990) tentatively classified it as WC based on their narrow-band photometry.
Our spectrogram (Figure 2) is unambiguous:
with very strong C~{\sc iv} lines at $\lambda$ 4650 and 5808
\AA\ , and weaker but still conspicuous lines of C~{\sc iii} $\lambda$ 5696
and O~{\sc iii}-{\sc v} $\lambda$ 5590, WR5 is a WC6 star,
according to the classification criteria of Crowther et al.
(1998).

One of the most intriguing stars in the sample is WR6, which
shows broad Balmer and HeI lines unlike most typical late-type WN stars.
An excellent spectrogram of this star, obtained nine years before
ours, is shown by Terlevich et al. (1996; see their Figure 7).
These authors call attention to the strong spectroscopic
variability of WR6's emission lines between 1980 and 1992. Our
spectrogram (Figure 4) is virtually identical to the one shown by Terlevich
et al., including line widths and He / H line ratios; this suggests a (temporary?) stagnation in the rapid spectroscopic evolution of this star observed recently. We must note that our spectrogram includes the strong continuuum from the star 0.6$''$ to the west of WR6, but our 2D spectrogram clearly shows that all the broad emission lines come
from the faintest of the pair.

Two WR stars, WR7 and WR11, are located in an area with high extinction,
according to the extinction map of Ma\'{\i}z-Apell{\'a}niz (2004; compare his Figure 6 with
our Figure 3). Their continuum and emission-line flux was therefore found to
be low in the previous imaging work. But the spectrum of WR7 is unambiguous
with very strong lines of C {\sc iv} at 4650 \AA\ and 5812 \AA\  but no evidence of
C {\sc iii} 5696, indicating a WC4
classification. WR11, first noticed as an emission-line star by DMS93, is not correctly identified in the original finding chart (their Figure 6). A careful superposition of the 1991 emission-line image from the Canada-France-Hawaii telescope, which was originaly used to identify WR candidates, with the more recent HST/ACS images, as well as an analysis of the location of the emission line in our more recent 2D spectra, clearly identifies WR11 as the brightest star in a very tight group of a half-dozen stars separated by less than 2$''$; its correct identification is now shown in Figure 3. The only emission feature visible in the spectrum of WR11 is a broad (FWHM = 37 \AA\  ) He {\sc ii} 4686 line. Because the group of stars in which WR11 stands is unresolved in ground-based spectrum, the true equivalent width of the line is certainly
much higher, making WR11 a genuine WN star. However, we 
find no evidence for He {\sc ii} 5411 nor C {\sc iv} 5800 in its
spectrum.

WR8 and WR10 were first detected in the CFHT images of DMS90, and Figure 1 shows
that their spectra are very similar. With well-resolved emission lines of He{\sc ii} 4686
(FWHM = 15 \AA ), N {\sc iii} 4634, 40 and C {\sc iv} 5808, as well as absorption features of N{\sc v} 4604 and 4620 \AA , the spectrum of these two stars is very similar to that of
the Galactic WN6 star HD 93162 (WR25; see Figure 1 in Walborn \& Fitzpatrick 2000).

WR12 was identified as having a weak He II excess by DMS93 
in their ground-based CFHT images, but not in their pre-CoStar
HST images because of the low S/N ratio. WFPC2 ultraviolet
(F170W filter) images shown by Bruhweiler et al. (2003)
clearly separate WR12 into two stars separated by less than
0.3$''$, labeled 690A (the faintest of the pair, classified as O5 III according to STIS UV
spectra) and 690B (B0 Ib). 
Based on this UV classification, one
should not expect to see He II 4686 in emission in either star (Walborn \&
Fitzpatrick 1990). Our slit
includes both stars, and the resulting spectrum (Figures 1 and 5) clearly shows a
broad emission bump (total W$_e$ = 15 \AA )
which includes the He II line (W$_e$ = 5
\AA\ confirming the early diagnostic based on imagery) accompanied 
by He{\sc i} 4713, N{\sc iii} 4634,40 and possibly C{\sc iv} 4650. We also detect
strong P Cygni profiles of He {\sc i} at 4388, 4471 and 4921 \AA\
as well as an absorption line of He {\sc ii} 5411; 
a careful examination of the 2D spectroscopic
images along the trace of WR12 clearly shows that these
lines are of stellar origin and are not significantly contaminated by nebular
emission.  The more extended spectrogram of WR12 shown in
Figure 5 bears striking similarities with that of the WN10 star Sk$-66^{o}40$. Although our optical spectrum is a composite
of two stars, it is more likely that most of the emission lines
come from the brightest one (690B). 

\begin{figure}
\includegraphics[width=84mm]{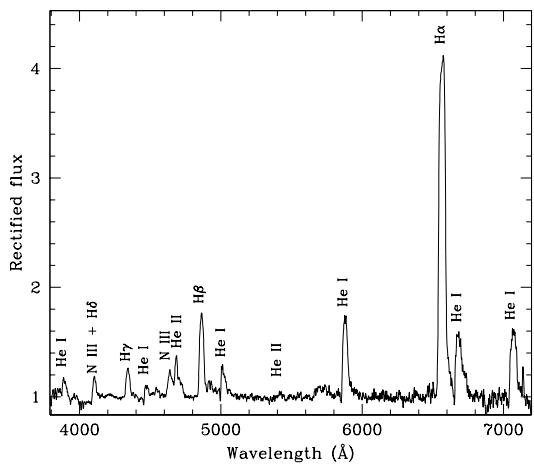}
\caption{Spectrum of NGC 604-WR6. The emission lines are diluted by the continuum from the bright star
0.6$''$ to the west of WR6. }
\end{figure}

\begin{figure}
\includegraphics[width=84mm]{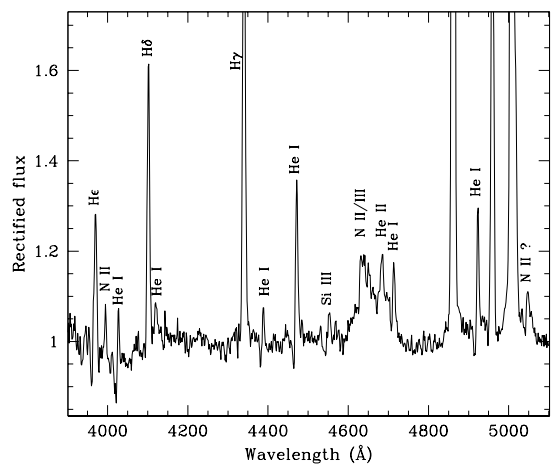}
\caption{Spectrum of NGC 604-WR12.}
\end{figure}

To confirm this, we have
analyzed WFPC2 images (program 5773) obtained with the nebular filters F656N
(H$\alpha$) and F673N (centered on the [SII] $\lambda\lambda$ 6717,31
doublet). Since stellar spectra do not show the presence of these
sulfur lines, and since the two nebular filters are close enough
in terms of wavelength (reducing possible extinction effects),
we have used the F673N image as a continuum reference (Figure 6) and subtracted it from
the H$\alpha$ image. Selected fields from the resulting
continuum-subtracted image are shown in Figures 7 and 8. Figure 7 demonstrates the power of this
technique on a field centered on NGC 604-WR6, which is known to be
a strong H$\alpha$ emitter (see Figure 4). Figure 8 shows the field around WR12. In addition to WR12, He {\sc ii} emission-line stars WR3, WR8 and V1 clearly show up as H$\alpha$ emitters. A comparison of the upper and lower panels of Figure 8
also confirms that, as expected, star 690B is a strong H$\alpha$ emitter whereas 690A does not
show up.	
	
\begin{figure}
\includegraphics[width=84mm]{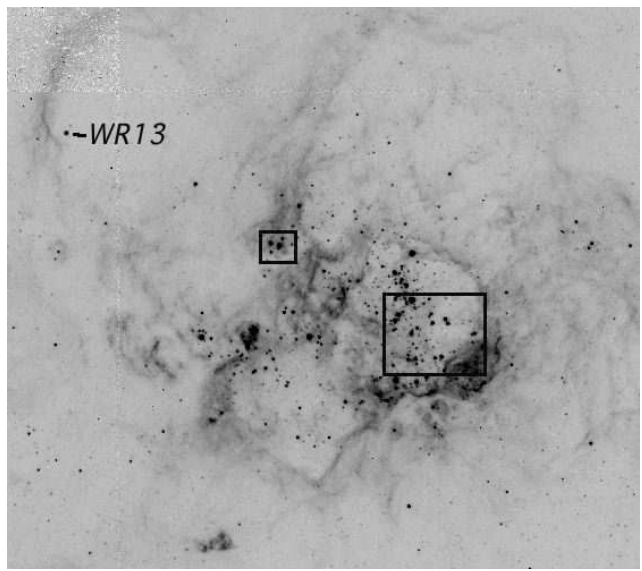}
\caption{WFPC2 F673N image of NGC 604, showing the location of WR13 and the two regions enlarged in Figures 7 and 8. North is at the top, east to the left ($30'' \times 30''$).}
\end{figure}

\begin{figure}
\includegraphics[width=84mm]{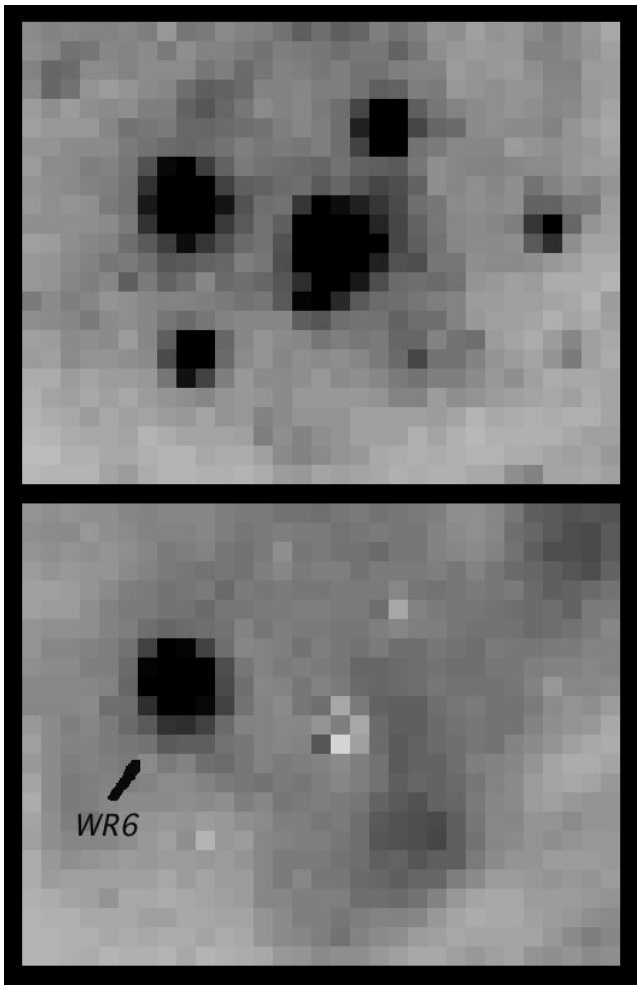}
\caption{WFPC2 images of the region surrounding NGC 604-WR6 in "continuum" (upper panel, F673N filter)
and in ``pure H$\alpha$" (lower panel; F656N - F673N). The field of view is 3.1$'' \times 2.4''$, with North up and East to the left.}
\end{figure}

\begin{figure}
\includegraphics[width=84mm]{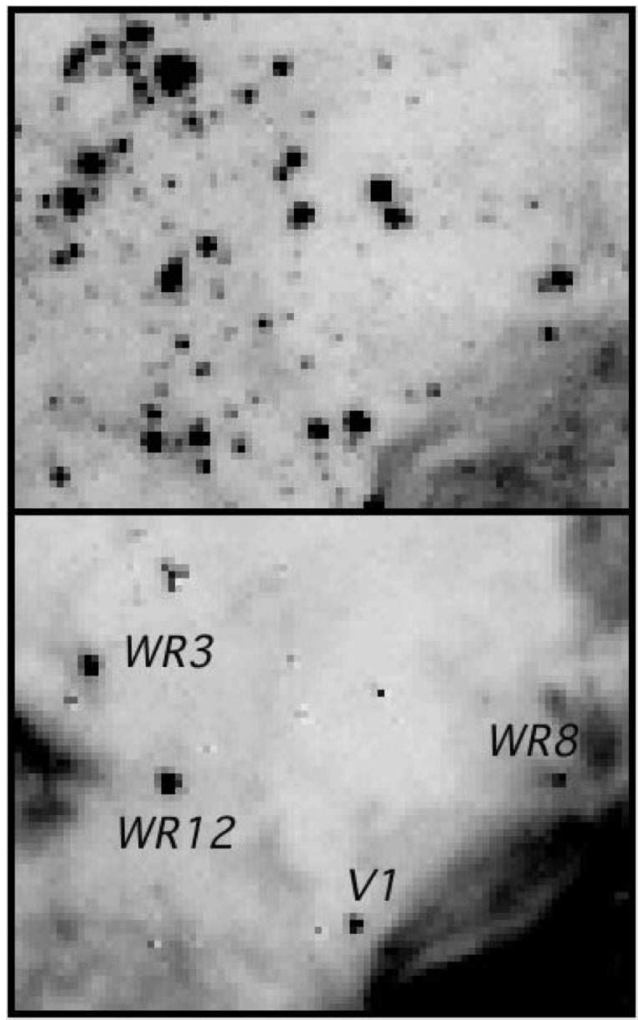}
\caption{WFPC2 images of the region surrounding NGC 604-WR3, WR12, WR8 and V1 in "continuum" (upper panel, F673N filter) and in ``pure H$\alpha$'' (F656N - F673N). Note the elongation of WR12 in the upper panel, corresponding to stars 690B (lower, brighter component: WR12 itself) and 690A (fainter, upper component) in
Bruhweiler et al. (2003). Only star 690B shows an excess of H$\alpha$. The field of view is $9.3'' \times 7.5''$, with North up and East to the left.}
\end{figure}

\begin{figure}
\includegraphics[width=84mm]{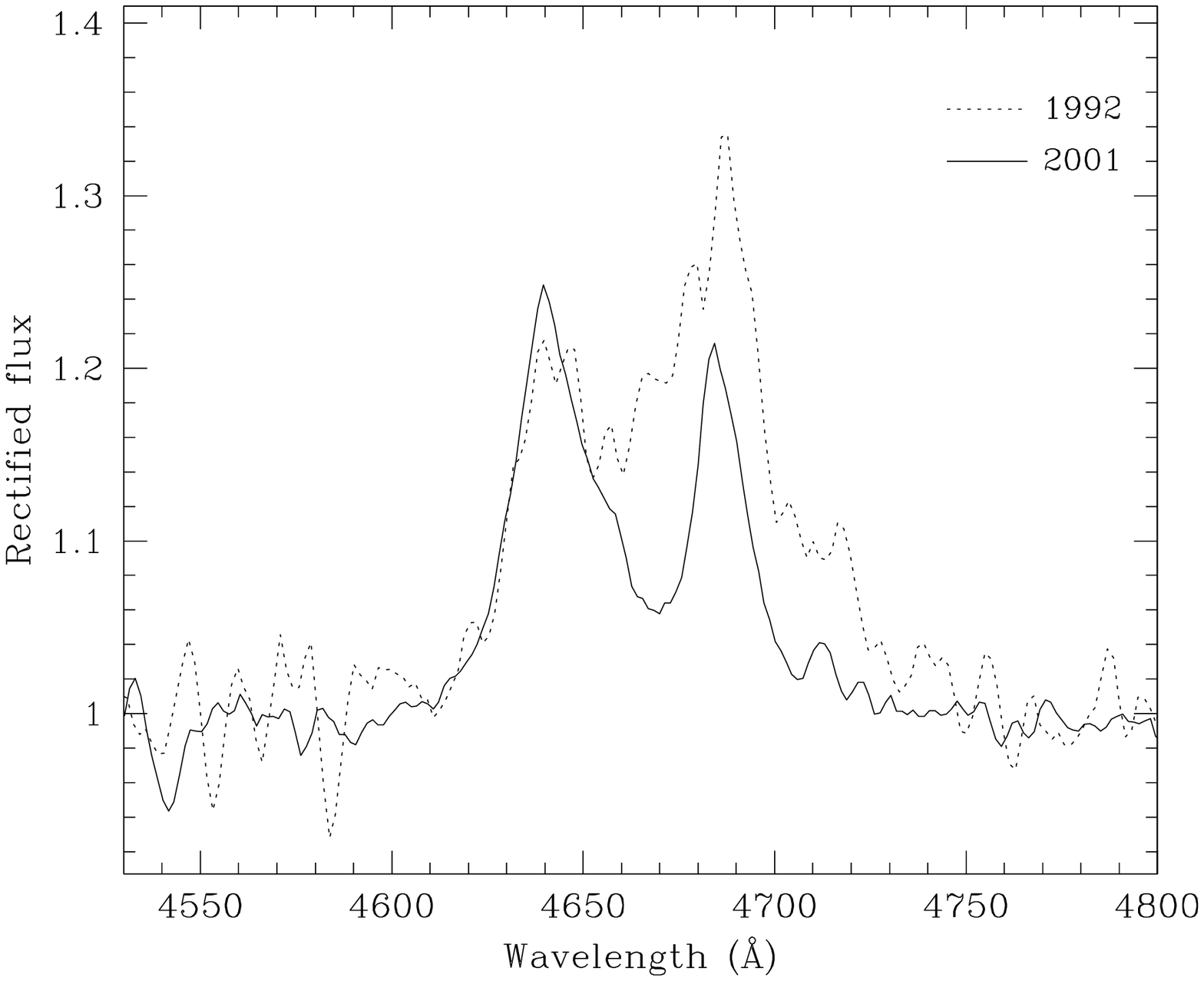}
\caption{Comparison between the 1992 and 2001 spectra of NGC 604 - V1.}
\end{figure}

N604-V1 was found to be photometrically variable (on a timescale of one night, possibly eclipsing?) by
Drissen, Moffat \& Shara (1990). DMS93 allude to a spectrogram, obtained in October 1992, showing He {\sc ii} in emission, but did not
publish it. The more recent spectrogram (October 2001) shows this line (FWHM = 14 \AA ), 
as well as N {\sc iii} 4640 in emission, absorption lines of He {\sc ii} 4541 and 5411,
and P Cygni profiles for the He {\sc i} lines
at 4471 and 5876 \AA . We also detect a weak (W$_{e}$ = 1.0 \AA ) emission line of C {\sc iii} $\lambda$ 5696.
A comparison between spectra obtained in
1992 and 2001 shows a significant weakening (by a factor of 2.5) of the He
{\sc ii} 4686 line (Figure 9). The 1992 spectrum did not show the He {\sc ii} absorption lines (perhaps filled
with emission), nor the C {\sc iii} $\lambda$ 5696 emission.

Finally, WR13 was never identified as a WR candidate but since it is the brightest star
inside a prominent ionized arc in the northeastern part of the nebula (see Figure 6), we took advantage of the
MOS capabilities and superposed a slitlet on it to satisfy our curiosity. 
Its spectrum shows very weak emission lines of He{\sc ii}
4686 (W$_e$ = 1.2 \AA ) and N{\sc iii} 4640 (W$_e$ = 3.3 \AA ),
as well as absorption lines of He{\sc ii} 4541 and 5411 and
He{\sc i} 4471. A comparison with stars in the atlas of Walborn
\& Fitzpatrick (1990) suggests a spectral type O6.5 Iaf, similar
to the galactic star HD 163758.

\subsection{NGC 595}

\begin{figure}
\includegraphics[width=84mm]{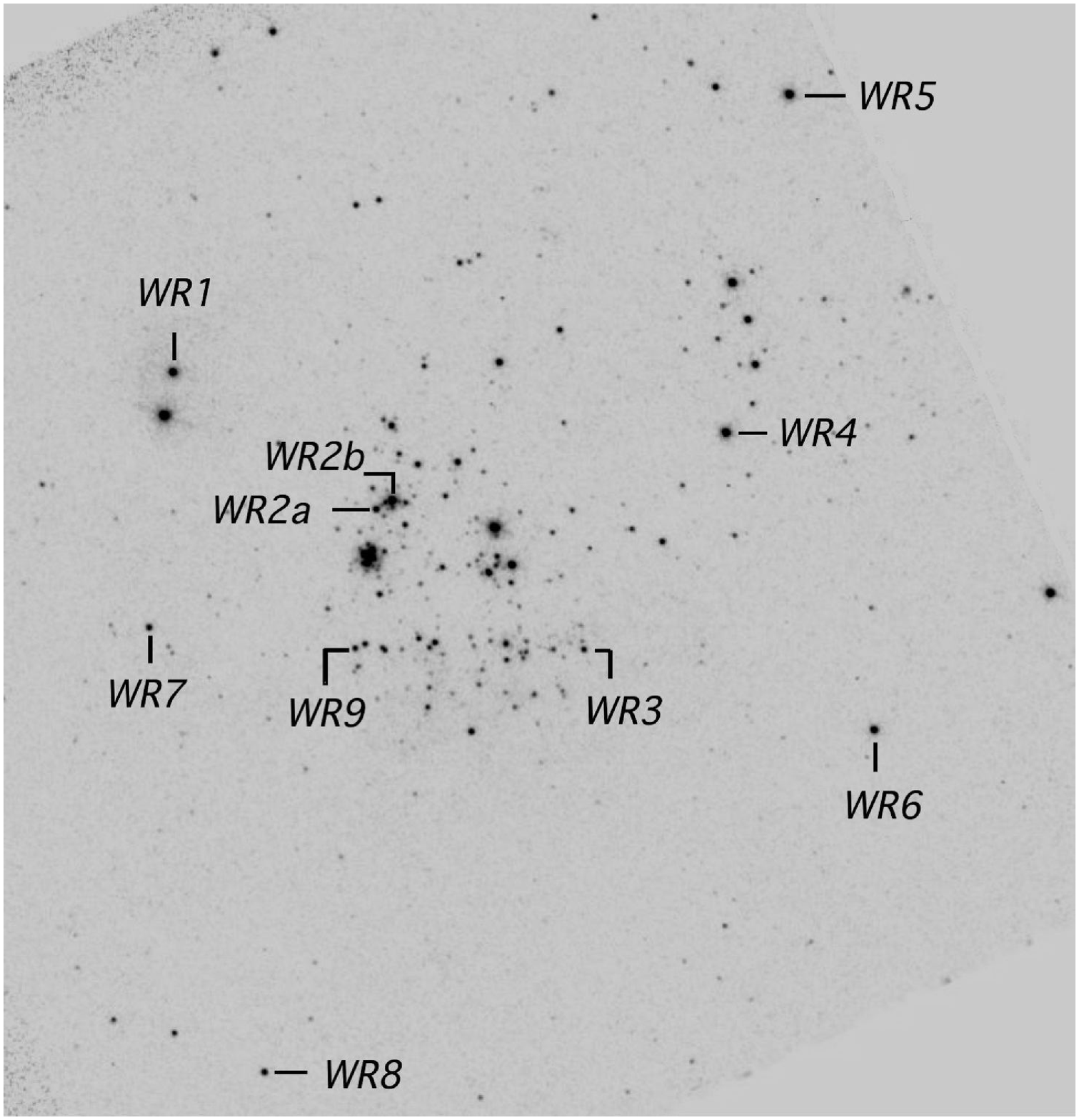}
\caption{Finding chart for emission-line stars in NGC 595 from HST/WFPC2 images (combination of F336W, F439W and F547M). The field of view is $30'' \times 35''$, with North at the top and East to the left. }
\end{figure}

An updated finding chart for W-R stars in NGC 595 is shown in Figure 10. 
Among all W-R candidates in NGC 595, only WR2 (components a and b), WR9 and WR11 lacked a clear, individual, spectroscopic identification, althought spectrograms including both WR2ab and WR11 (both objects, separated by less than 2$''$, are themselves multiple as shown by WFPC2 images) as well as WR9 clearly show the presence of at least one WNL star (Conti \& Massey 1981, Massey et al. 1996). But a re-examination of the HST/WFPC images discussed by DMS93 shows that the excess of light at 4686 \AA\ at the location of the candidate W-R star WR11 is marginal at best, and no emission excess could be detected for WR10. Indeed, Massey \& Johnson (1998) question the presence of emission lines in WR10.
Moreover, Royer et al. (2003) unambiguously detect WR2 and WR9 as W-R candidates in their images obtained with a set of narrow-band filters, but also fail to detect WR10 and WR11. So we decided to remove both objects from our list of W-R candidates and therefore they do not appear in Figure 10.

The only new spectroscopic confirmation from this paper is that of NGC 595-WR9. Its spectrum (Figure 1) is comparable to that of the Galactic star WR22 (HD 92740; see Figure 1 in Walborn \& Fitzpatrick 2000), with W$_{e}$ = 18 \AA\ for He {\sc ii} 4686 and 9 \AA\ for N {\sc iii} 4634, 40. We also detect an emission line of He {\sc ii} 5411 (W$_{e}$ = 4.5 \AA ).

The first spectrogram of WR3 (MC29), published by Massey \& Conti (1983), allowed a crude WC classification. Line ratios (C {\sc iv} 5808, C {\sc iii} 5696 and O {\sc iii,v} 5590) in our spectrogram (Figure 2) allows us to refine the classification to WC6. It is interesting to note that WR3 is the only WC star in NGC 595.

NGC 595-WR1 and NGC 595-WR5 have very similar spectra in terms of line ratio, and both can be classified as WN7. The equivalent width of all lines are larger by a factor of two for WR5; this could be due, at least in part, to some contamination by the continuum of the bright star 1.3$''$ south of WR1. NGC595-WR5 is in a region of relatively weak nebular emission, and we were able to subtract the nebular component from its spectrum, which is shown in its entirety in Figure 11. The relative strengths of the pure helium lines (4200 \AA , 4541 \AA , and 5411 \AA ) to those for which helium and hydrogen both contribute (4340 \AA\ and 4860 \AA ) clearly indicate the presence of hydrogen in the atmosphere of NGC 595-WR5.

\begin{figure}
\includegraphics[width=84mm]{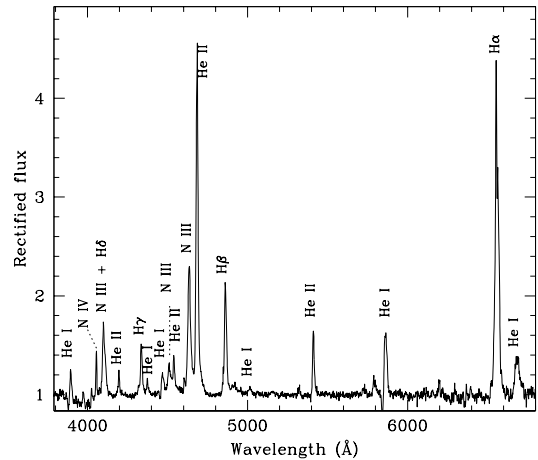}
\caption{Spectrum of NGC 595-WR5. The (weak) surrounding nebular component has been subtracted, leaving only stellar lines.}
\end{figure}

\subsection{NGC 592 and NGC 588}

\begin{figure}
\includegraphics[width=84mm]{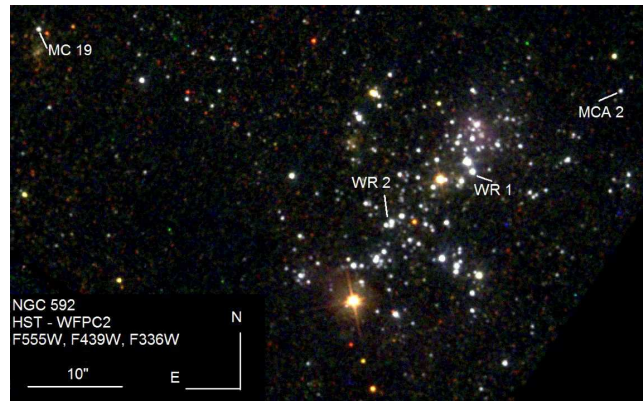}
\caption{Composite visible image of NGC 592 from HST/WFPC2}
\end{figure}

NGC 592 and NGC 588 are much less luminous and rich in WR stars than NGC 595 and NGC 604, by factors of 3 to 10, but they nevertheless include a fair population of young, massive stars.
We defer the analysis of their stellar content, based on HST archives, to another paper (\'Ubeda, Drissen \& Crowther, in preparation), but we present here a spectrum and some images relevant to their emission-line star content.

NGC 592 is ten times less luminous
than NGC 604, both in terms of H$\alpha$ luminosity 
(Bosch et al. 2002) and in UV continuum (as measured
with the large FUSE aperture at 1150~\AA; see Pellerin 2006), 
and until now  very little is known about its stellar content. 
It includes four WR stars: two in the core (identified as WR1 and WR2 
following DMS90), and two more in the outskirts: MC19 and MCA2. Their positions  are labeled in a composite
visible image in Figure 12 as WR1, WR2, MC19 and MCA2.
Despite the factor ten smaller in H$\alpha$ luminosity compared to NGC 604, Figure 12 shows that the ionizing cluster of NGC 592 still contains a fair number of massive stars. 

A spectrogram of NGC~592--WR1 (WNL)
is  shown in Conti \& Massey (1981, CM3).
NGC~592--WR1  (also known as MC17,
or WR25 in Massey \& Johnson 1998) was originally misidentified with the brighter
but slightly redder companion 1$''$ to the north-east 
(star B67 in Humphreys \& Sandage 1980), but it was already  
strongly suggested by the images presented in Drissen et al. (1990) that the WR star 
was not the brightest component of the tight pair; it is 
now obvious in the WFPC2 images. NGC~592--WR1 is also the cluster's brightest UV source 
in the F170W image.

NGC~592--WR2, first identified as a WR candidate by Drissen et al. (1990),  lacked spectral confirmation 
until now. Our spectrogram of this star is
shown in Figure 13, and is that of a WN4 star as deduced from the weakness of 
the He~{\sc i} 5876 line and the absence of N~{\sc iii} 4640. A comparison of our spectrum 
with those of a dozen Galactic WNE stars shows that the red spectrum of NGC~592--WR2  
most closely resembles that of the Galactic WN4 star WR44, 
analysed by Hamann \& Koesterke (1998).
A careful
comparison of the WFPC2 image and the CFHT 2D spectra clearly
shows that WR2 is the brightest of a close pair separated by
0.6$''$.
                           
The spectrum of MC19 (WCE) was first published by
Massey \& Conti (1983), while MCA2 is a WN star whose spectrum has been published by  Massey, Conti \& Armandroff (1987). 

\begin{figure}
\includegraphics[width=84mm]{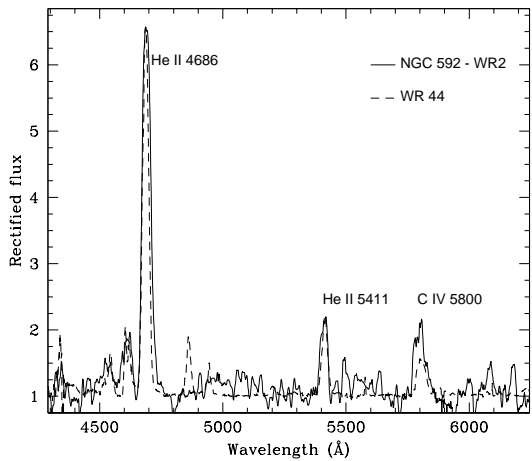}
\caption{Full spectrogram of NGC 592-WR2, compared with that of
the Galactic WN4 star WR44 (LSS 2289). Note that all nebular
lines have been removed from the spectrogram of NGC 592 - WR2.}
\end{figure}

NGC 588 was not part of the imaging survey by Drissen, Moffat \& Shara (1990). 
However, two emission-line stars are known in this cluster: MC3, a WNL, detected with narrow-band imagery by Conti \& Massey (1981) and spectroscopically confirmed by Massey \& Conti (1983); and UIT-008 , a transition Of/WN9 star (see  Massey \& Johnson 1998 for a visible spectrogram and Bianchi, Bohlin \& Massey 2004 for a UV spectrogram) selected for follow-up spectroscopy because of its high UV luminosity. NGC 588 is, like the other giant HII regions discussed in this paper, a crowded place and high-resolution images are required to properly identify the stars. Fortunately, NGC 588 was imaged with WFPC2 in many bands in 1994 (Figure 14), including F469N, 
a narrowband filter centered on the He{\sc ii} 4686 emission line. Only MC3 clearly stands out in the F469N image; 
 it is 0.5 mag brighter than in the F439W image. UIT-008 is marginally brighter in the F469N image than in the continuum F439W, by 0.1 mag; this is consistent with the weakness of its emission lines.
 
 \begin{figure}
\includegraphics[width=84mm]{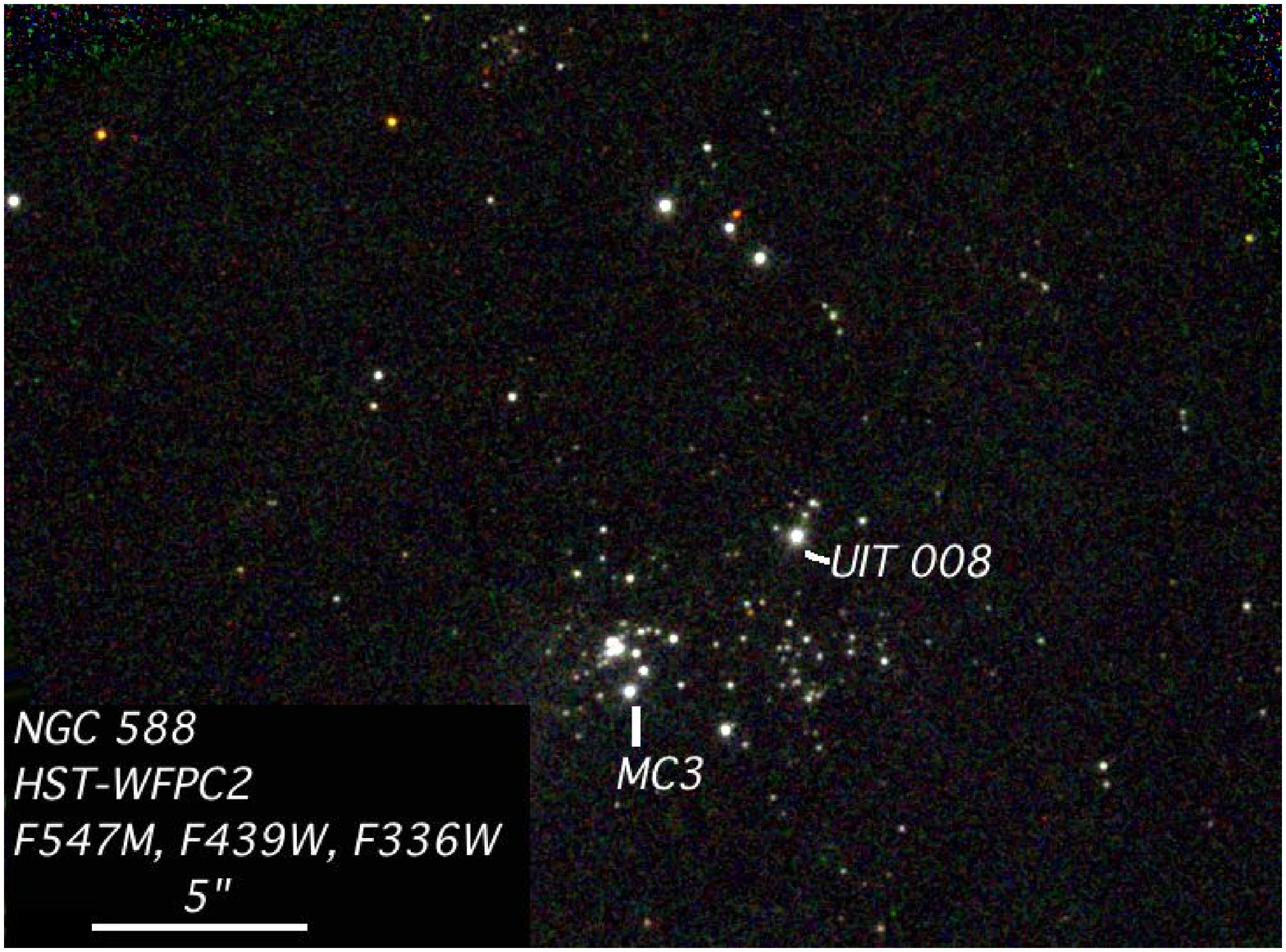}
\caption{Composite visible image of NGC 588 from HST/WFPC2.}
\end{figure}

 \section{Conclusions}
We have spectroscopically confirmed the presence of emission lines in a sample of WR candidates selected by interference filter imagery. Most of them are genuine WR stars, but we have also detected transition-type evolved O stars, with an He II equivalent width as low as 5 \AA\ . As demontrated here and in previous publications, this technique is very efficient, especially in crowded fields (see also Hadfield \& Crowther 2007), with a very high success rate and relatively few false detections. The census of the WR population in NGC 588, NGC 592, NGC 595 and NGC 604 is now essentially complete: only three stars (NGC 604-WR2, NGC 604-WR4 and NGC 595-WR2), which are members of very dense and barely resolved groups, lack a clear identification although their WR nature is not in doubt.
We have also obtained good quality spectrograms of previously known WR stars in these regions, allowing a better spectral type identification. The four giant HII region studied here harbour about 20\% of the entire known Wolf-Rayet population of M33, with a WC/WN ratio significantly lower (0.25) than that of the field (0.4; Massey \& Johnson 1998).

\section*{Acknowledgements}

We thank the referee for a carefull reading of the manuscipt and suggestions.
This paper is based on observations obtained at the Canada-France-Hawaii Telescope (CFHT) which is operated by the National Research Council of Canada, the Institut National des Sciences de l'Univers of the Centre National de la Recherche Scientifique of France,  and the University of Hawaii.
LD is grateful to the Canada Research Chair program, the Natural Sciences and Engineering Research Council of Canada, and the Fonds Qu\'eb\'ecois de la Recherche sur la Nature et les Technologies (FQRNT) for financial support. LU acknowledges FQRNT for a postdoctoral fellowship.
PAC wishes to thank the Royal Society for providing financial assistance through their wonderful University Research Fellowship scheme for the past eight years.

\clearpage

\label{lastpage}


\begin{thebibliography}{99}

\bibitem[\protect\citeauthoryear{Abbott et al.}{2004}]{2004MNRAS.350..552A} 
Abbott J.~B., Crowther P.~A., Drissen L., Dessart L., Martin P. \& Boivin G., 2004, MNRAS, 350, 552 

\bibitem[\protect\citeauthoryear{Armandroff \& Massey}{1985}]{1985ApJ...291..685A} Armandroff T.~E., \& Massey P., 1985, ApJ, 
291, 685 

\bibitem[\protect\citeauthoryear{Armandroff \& 
Massey}{1991}]{1991AJ....102..927A} Armandroff T.~E., \& Massey P., 1991, AJ, 102, 927 

\bibitem[\protect\citeauthoryear{Bianchi}{2004}]{b1} 
Bianchi, L., Bohlin, R., \& Massey, P. 2004, ApJ, 601, 228

\bibitem[\protect\citeauthoryear{Bosch, Terlevich, \& 
Terlevich}{2002}]{2002MNRAS.329..481B} Bosch G., Terlevich E., \& Terlevich 
R., 2002, MNRAS, 329, 481

\bibitem[\protect\citeauthoryear{Bruhweiler, Miskey, \& Smith 
Neubig}{2003}]{2003AJ....125.3082B} Bruhweiler F.~C., Miskey C.~L., \& Smith 
Neubig M., 2003, AJ, 125, 3082

\bibitem[\protect\citeauthoryear{Conti \& Massey}{1981}]{1981ApJ...249..471C} Conti P.~S., \& Massey P., 1981, ApJ, 249, 471 

\bibitem[\protect\citeauthoryear{Crowther, de Marco, \& Barlow}{1998}]{1998MNRAS.296..367C} Crowther P.~A., de Marco O., \& Barlow 
M.~J., 1998, MNRAS, 296, 367 

\bibitem[\protect\citeauthoryear{D'Odorico \& 
Rosa}{1981}]{1981ApJ...248.1015D} D'Odorico S., \& Rosa M., 1981, ApJ, 248, 
1015 

\bibitem[\protect\citeauthoryear{Drissen, Moffat, \& 
Shara}{1990}]{1990ApJ...364..496D} Drissen L., Moffat A.~F.~J., \& Shara 
M.~M., 1990, ApJ, 364, 496 

\bibitem[\protect\citeauthoryear{Drissen, Moffat, \& 
Shara}{1993}]{1993AJ....105.1400D} Drissen L., Moffat A.~F.~J., \& Shara 
M.~M., 1993, AJ, 105, 1400 

\bibitem[\protect\citeauthoryear{Hadfield}{2007}]{Hadfield07} 
Hadfield, L. J., \& Crowther, P. A. 2007, MNRAS, 381, 418

\bibitem[\protect\citeauthoryear{Hamann}{1998}]{Hamann98} 
Hamann, W.-R., \& Koesterke, L. 1998, A\&A, 333, 251

\bibitem[\protect\citeauthoryear{Holtzman et al}{1995}]{Holtzman95} Holtzman, J. A., Burrows, C. J., Casertano, S., Hester, J. J., Trauger, J. T., Watson, A. M., \& Worthey, G. 1995, PASP, 107, 1065

\bibitem[\protect\citeauthoryear{Humphreys \& 
Sandage}{1980}]{1980ApJS...44..319H} Humphreys R.~M., \& Sandage A., 1980, ApJS, 44, 319 

\bibitem[\protect\citeauthoryear{Magrini et al.}{2007}]{} Magrini, L., V\'ilchez, J. M., Mampaso, A., Corradi, R. L. M., \& Leisy, P. 2007, A\&A, 470, 865 
 

\bibitem[Maiz(2004)]{Jesus04} Ma\'{\i}z-Apell\'aniz, J., P\'erez, E., \& Mas-Hesse, J. M.  2004, AJ, 128, 1196

  
\bibitem[\protect\citeauthoryear{Massey}{1996}]{b11} Massey, P., Bianchi, L., Hutchings, J.~B.,  \& Stecher, T.~P. 1996, ApJ, 469, 629

\bibitem[\protect\citeauthoryear{Massey \& Conti}{1983}]{1983ApJ...273..576M} Massey P., \& Conti P.~S., 1983, ApJ, 273, 576

\bibitem[\protect\citeauthoryear{Massey \& Johnson}{1998}]{1998ApJ...505..793M} 
Massey P., \& Johnson O., 1998, ApJ, 505, 793 

\bibitem[\protect\citeauthoryear{Massey et al.}{1987}]{Massetal87} 
Massey, P., Conti, P. S., \& Armandroff, T. E. 1987, AJ, 94, 1538
   
\bibitem[\protect\citeauthoryear{Meynet \& Maeder}{2005}]{2005A&A...429..581M} Meynet G., \& Maeder A., 2005, A\&A, 429, 581 

\bibitem[\protect\citeauthoryear{Pellerin}{2006}]{2006AJ....131..849P} Pellerin A., 2006, AJ, 131, 849 

 
\bibitem[\protect\citeauthoryear{Pindao et al.}{2002}]{2002A&A...394..443P} 
Pindao M., Schaerer D., Gonz{\'a}lez Delgado R.~M., \& Stasi{\'n}ska G., 2002, 
A\&A, 394, 443

\bibitem[\protect\citeauthoryear{Royer, Lundstr{\"o}m, \& 
Vreux}{2003}]{2003IAUS..212..572R} Royer P., Lundstr{\"o}m I., \& Vreux J.-M., 
2003, in A Massive Star Odyssey: From Main Sequence to Supernova, Proceedings of IAU Symposium 212, ed.  
K. van der Hucht, A. Herrero, C. Esteban (San Francisco: Astronomical Society of the Pacific), 572

\bibitem[\protect\citeauthoryear{Terlevich et al.}{1996}]{1996MNRAS.279.1219T} Terlevich E., 
D\'iaz A.~I., Terlevich R., Gonz\'alez-Delgado R.~M., P\'erez E., \& Garc\'ia Vargas M.~L., 1996, 
MNRAS, 279, 1219

\bibitem[\protect\citeauthoryear{Vacca et al.}{1995}]{1995ApJ...444..647V} 
Vacca W.~D., Robert C., Leitherer C., \& Conti P.~S., 1995, ApJ, 444, 647

\bibitem[\protect\citeauthoryear{Walborn \& 
Fitzpatrick}{1990}]{1990PASP..102..379W} Walborn N.~R., \& Fitzpatrick E.~L., 
1990, PASP, 102, 379 

\bibitem[\protect\citeauthoryear{Walborn \& 
Fitzpatrick}{2000}]{2000PASP..112...50W} Walborn N.~R., \& Fitzpatrick E.~L., 
2000, PASP, 112, 50

\end{thebibliography}
\end{document}